\documentclass[reprint,aps,pra,showpacs,superscriptaddress]{revtex4-1} 
\usepackage{amsmath,amssymb}
\usepackage{amsgen,amsfonts,amsbsy}
\usepackage{mathrsfs}
\usepackage{bbm}
\usepackage[dvipdfmx]{graphicx, color}
\usepackage{dcolumn}
\usepackage{bm}
\usepackage{enumerate}
\usepackage{theorem}

\def\tr{\mathop{\mathrm{tr}}}
\def\Tr{\mathop{\mathrm{Tr}}}

\begin{document}

\title{
Effect of nonnegativity on estimation errors \\
in one-qubit state tomography with finite data
}

\author{Takanori Sugiyama}
\email{sugiyama@eve.phys.s.u-tokyo.ac.jp}
\affiliation{
Department of Physics, Graduate School of Science, The University of Tokyo,\\
7-3-1 Hongo, Bunkyo-ku, Tokyo, Japan 113-0033.
}
\author{Peter S. Turner}
\email{turner@phys.s.u-tokyo.ac.jp}
\affiliation{
Department of Physics, Graduate School of Science, The University of Tokyo,\\
7-3-1 Hongo, Bunkyo-ku, Tokyo, Japan 113-0033.
}
\author{Mio Murao}
\email{murao@phys.s.u-tokyo.ac.jp}
\affiliation{
Department of Physics, Graduate School of Science, The University of Tokyo,\\
7-3-1 Hongo, Bunkyo-ku, Tokyo, Japan 113-0033.
}
\affiliation{
Institute for Nano Quantum Information Electronics, The University of Tokyo,\\
4-6-1 Komaba, Meguro-ku, Tokyo, Japan 153-8505.
}

\date{\today}

\begin{abstract}
   We analyze the behavior of estimation errors evaluated by two loss functions, the Hilbert-Schmidt distance and infidelity, in one-qubit state tomography with finite data. 
   We show numerically that there can be a large gap between the estimation errors and those predicted by an asymptotic analysis.
   The origin of this discrepancy is the existence of the boundary in the state space imposed by the requirement that density matrices be nonnegative (positive semidefinite).
   We derive an explicit form of a function reproducing the behavior of the estimation errors with high accuracy by introducing two approximations: a Gaussian approximation of the multinomial distributions of outcomes, and linearizing the boundary.
   This function gives us an intuition for the behavior of the expected losses for finite data sets.
   We show that this function can be used to determine the amount of data necessary for the estimation to be treated reliably with the asymptotic theory.
   We give an explicit expression for this amount, which exhibits strong sensitivity to the true quantum state as well as the choice of measurement.  
\end{abstract}

\maketitle

\section{Introduction}\label{sec:intro}

Quantum tomography has become a standard measurement technique in quantum physics. 
It is especially important in the field of quantum information as it is used for the confirmation of successful experimental implementation of quantum protocols.
For example, it can be used to confirm that the quantum states required in a quantum information protocol are sufficiently closed to their theoretical targets \cite{QTText2004}.
In practice, experimental data obtained from tomographic measurements are used to assign a mathematical description to an unknown quantum state or operation, called an estimate.   Statistically, this is a constrained multi-parameter estimation problem -- the quantum estimation problem -- where we assume we are given a finite number of identical copies of a quantum state or process, we perform measurements whose mathematical description is assumed to be known, and from the outcome statistics we make our estimate.
Due to the probabilistic behavior of the measurement outcomes and the finiteness of the number of measurement trials, there always exist statistical errors in any quantum estimate. 
The size of the error depends on the choice of the measurements, known as the experimental design, and the estimation algorithm, known as the estimator.
A standard combination in quantum information is that of quantum tomography and maximum likelihood estimation \cite{QTText2004}.  
In order to compare estimation schemes, it is therefore important to evaluate precisely the size of the estimation error for a given combination of experimental design and estimator.

For evaluating the size of the estimation error, we introduce a distance-like function, called a loss function,  between the estimate and the true operator.
One way to evaluate estimation errors using a loss function is an expected loss, which is the statistical expectation value of the loss function over all possible data sets. 
In quantum information experiments, the infidelity (one minus the fidelity) and the trace distance are often used as loss functions for state estimation.
These evaluations are often performed in the theoretical limit of infinite data, called the asymptotic regime.
The asymptotic behavior of these expected losses for this combination has been studied very well \cite{GillMassar2000, HayashiSelectedPapers2005}.
Using the asymptotic theory of parameter estimation, we can show that for a sufficiently large number of measurement trials, $N$, there is a lower bound of the expected losses, called the Cram\'{e}r-Rao bound.
 It is known that a maximum likelihood estimator achieves the Cram\'{e}r-Rao bound asymptotically, and that those expected losses decrease as $O(1/N)$.

In practice of course, no experiment produces infinitely many data, and there are problems in applying the asymptotic theory of expected losses to finite data sets. 
First of all, the Cram\'{e}r-Rao inequality holds only for a specific class of estimators, namely those that are unbiased.
A maximum likelihood estimator is asymptotically unbiased, but is not unbiased for finite $N$, so the expected losses can be smaller than the bound for finite $N$. 
Especially when the purity of the true density matrix becomes high, the bias becomes larger.
This is due to the boundary in the parameter space imposed by the condition that density matrices be positive semidefinite, and the expected losses can deviate significantly from the asymptotic behavior \cite{Burgh2008, Sugiyama2012_1}.
A natural question is then to ask at what value of $N$ the expected losses begin to behave asymptotically.
If $N$ is large enough for the effect of the bias to be negligible, we can safely apply the asymptotic theory for evaluating the estimation error in an experiment.
However, in general, determining the effects of the bias is a difficult problem.
 
%

In this paper, we analyze the effect of the bias caused by the parameter space boundary in one-qubit state tomography using a maximum likelihood estimator.  
In section \ref{sec:review}, we briefly review quantum state tomography and the asymptotic theory.
In section \ref{sec:theory}, we analyze the boundary effect theoretically.
Applying ideas from classical statistical estimation theory, we derive an approximate form of the expected losses for finite $N$.
In section \ref{sec:Numerical}, we analyze the boundary effect numerically, giving the results of our pseudo-random numerical experiments.
These indicate that the function we derived reproduces the behavior of the expected losses for finite $N$ more precisely than the Cram\'{e}r-Rao bound. 
This makes it possible to predict the point at which the behavior of the expected infidelity becomes effectively asymptotic.
We conclude in section \ref{sec:Conclusion}.

\section{Quantum state tomography and asymptotic estimation theory}\label{sec:review}

   In this section, we give a brief review of known results in quantum state tomography and asymptotic estimation theory.
   The purpose of quantum state tomography is to identify the density matrix characterizing the state of a quantum system of interest.   
   Here we only consider states of a single qubit.
   Let $\mathcal{H}$ be the 2-dimensional Hilbert space $\mathbb{C}^2$ and $\mathcal{S}(\mathbb{C}^2)$ be the set of all positive semidefinite density matrices acting on $\mathcal{H}$.
  Such a density matrix $\rho$ can be parametrized as 
  \begin{eqnarray}
     \rho (\bm{s}) = \frac{1}{2}(\mathbbm{1} + \bm{s} \cdot \bm{\sigma}),
  \end{eqnarray}
  where $\mathbbm{1}$ is the identity matrix on $\mathbb{C}^2$, $\bm{\sigma} = (\sigma_{1}, \sigma_{2}, \sigma_{3})^\mathrm{T}$ is the vector of Pauli matrices, and $\bm{s} \in \mathbb{R}^3,\ \| \bm{s} \| \le 1$, is called the Bloch vector. 
   Let us define the parameter space $S:=\{ \bm{s}|\ \rho (\bm{s})\in \mathcal{S}(\mathbb{C}^2 )\}$.
   Identifying the true density matrix $\rho \in \mathcal{S}(\mathbb{C}^{2})$ is equivalent to identifying the true parameter $\bm{s} \in S$. 
   Let $\bm{\Pi}=\{ \Pi_{x} \}_{x\in \mathcal{X}}$ denote the POVM characterizing the measurement apparatus used in the tomographic experiment, 
   where $\mathcal{X}$ is the set of measurement outcomes.
   Like a density matrix, a POVM can be parametrized as 
   \begin{eqnarray}
      \Pi_{x} = v_{x}\mathbbm{1} + \bm{w}_{x} \cdot \bm{\sigma},
   \end{eqnarray}
   where $(v_x, \bm{w}_x) \in \mathbb{R}^4$.
   When the true density matrix is $\rho (\bm{s})$, Born's Rule tells us that the probability distribution describing the tomographic experiment is given by 
   \begin{eqnarray}
      p(x | \bm{s} ) &= \mbox{Tr}[\rho(\bm{s})\Pi_{x}]\\
                             &= v_{x} + \bm{w}_{x} \cdot \bm{s},
   \end{eqnarray} 
   where $\Tr$ denotes the trace operation with respect to $\mathbb{C}^2$.
   We assume that in the experiment we prepare identical copies of an unknown state $\rho (\bm{s})$.
   We perform $N$ measurement trials and obtain a data set $\bm{x}^N = (x_{1}, \ldots , x_{N})$, where $x_{i} \in \mathcal{X}$ is the outcome observed in the $i$-th trial.
   Let $N_{x}$ denote the number of times that outcome $x$ occurs in $\bm{x}^N$, then $f_{N}(x):=N_{x}/N$ is the relative frequency of $x$ for the data set $\bm{x}^N$.
   In the relative frequency interpretation of probability, one has that in the limit of $N \to \infty$, $f_{N}(x)$ converges to the true probability $p(x|\bm{s})$. 
   A POVM is called \emph{informationally complete} if $\Tr [\rho \Pi_{x} ] = \Tr [ \rho^{\prime} {\Pi}_{x}]$ has a unique solution $\rho^{\prime}$ for arbitrary $\rho \in \mathcal{S}(\mathcal{H})$ \cite{Prugovecki77}.
   This condition is equivalent to that of the POVM $\bm{\Pi}$ being a basis for the set of all Hermitian matrices on $\mathcal{H}$.
   For finite $N$, the relative frequency and true probability are generally not the same, {\it i.e.}, there is unavoidable statistical error, and we need to choose an estimation procedure that takes the experimental result $\bm{x}^N$ to a density matrix, that is, we need an estimator.   

   It is natural to consider a linear estimator, which demands that we find a $2 \times 2$ matrix $\rho^{\mathrm{li}}_{N}$ satisfying 
   \begin{eqnarray}
      \Tr [ \rho^{\mathrm{li}}_{N} \Pi_{x} ] = f_{N}(x),\ x\in \mathcal{X}. \label{eq:Linear}
   \end{eqnarray} 
   However, Eq.(\ref{eq:Linear}) does not always have a solution, and even when it does, although the solution is Hermitian and normalized, it is not guaranteed that $\rho^{\mathrm{li}}_{N}$ is positive semidefinite. 
   Let us explore this point further in the one qubit case.
   The positive semidefinite condition restricts the physically permitted parameter region to the ball $B := \{  \bm{s} \in \mathbbm{R}^{3}  | \| \bm{s}\| \le 1 \}$.
   On the other hand, a linear estimate is a random variable that can take values anywhere in the cube $C:= \{  \bm{s} \in \mathbbm{R}^3 | -1 \le s_{\alpha} \le 1, \alpha = 1,2,3  \}$. 
   There is therefore a `gap' between $B$ and $C$, consisting of unphysical linear estimates.
   When the true Bloch parameter $\bm{s}$ is in the interior of $B$ and $N$ becomes sufficiently large, the probability that linear estimates are out of $B$ becomes negligibly small. However, when the Bloch vector is on the boundary of $B$, or when $N$ is not sufficiently large, the effect of unphysical linear estimates cannot be ignored.      
   A maximum likelihood estimator $\rho^{\mathrm{ml}}$ is one way to address these problems.
   The estimated density matrix and the Bloch vector are defined as
   \begin{eqnarray}
      \rho^{\mathrm{ml}}_{N} &:= \mbox{argmax}_{\rho \in \mathcal{S}(\mathcal{H})} \prod_{i=1}^{N}\mbox{Tr}[ \rho {\Pi}_{x_i}], \\
      \bm{s}^{\mathrm{ml}}_{N} &:= \mbox{argmax}_{\bm{s} \in B} \prod_{i=1}^{N}\mbox{Tr}[ \rho (\bm{s}) {\Pi}_{x_i}].
   \end{eqnarray}
   It can be shown that when $ \rho^{\mathrm{li}}_{N} \in \mathcal{S}(\mathcal{H})$,
   $\rho^{\mathrm{li}}_{N} =\rho^{\mathrm{ml}}_{N}$ holds \cite{Hradil1997}.

   In order to evaluate the precision of estimates, we introduce a loss function.
   A loss function $\Delta$ is a map from $\mathcal{S} (\mathcal{H}) \times \mathcal{S} (\mathcal{H})$ to $\mathbb{R}$ such that (i) $\forall \rho, \sigma\in \mathcal{S}(\mathcal{H})$, $\Delta (\rho ,\sigma) \ge 0$, and (ii) $\forall \rho \in \mathcal{O}, \Delta (\rho , \rho ) = 0$.
   For example, the trace-distance and the infidelity (one minus the fidelity) are loss functions for density matrices.
     For our loss functions, we use both the squared Hilbert-Schmidt distance $\Delta^{\mathrm{HS}}$ and the infidelity $\Delta^{\mathrm{IF}}$ \cite{Bagan2004} defined as 
  \begin{eqnarray}
     \Delta^{\mathrm{HS}} (\bm{s}, \bm{s}^{\prime}) 
     &:=& \frac{1}{2} \Tr \left[ \bigl( \rho (\bm{s}) - \rho (\bm{s}^{\prime}) \bigr)^2 \right] \\
      &\phantom{:}=& \frac{1}{4} (\bm{s} - \bm{s}^{\prime} )^2 , \label{eq:HSdistance2_para}\\
     \Delta^{\mathrm{IF}} (\bm{s}, \bm{s}^{\prime} ) 
     &:=& 1 - \Tr \left[ \sqrt{ \sqrt{\rho (\bm{s})} \rho (\bm{s}^{\prime}) \sqrt{\rho (\bm{s}) } } \right]^2 \\
      &\phantom{:}=& \frac{1}{2}\left( 1 - \bm{s}\cdot \bm{s}^{\prime} - \sqrt{1-\|\bm{s}\|^2}\sqrt{1-\|\bm{s}^{\prime}\|^2 }  \right).\label{eq:Infidelity_para}
  \end{eqnarray} 
  The Hilbert-Schmidt distance is a normalized Euclidean distance in the parameter space, and the infidelity is a conventional loss function used in experiments. 
  We note that the Hilbert-Schmidt distance coincides with the trace distance in one-qubit systems, but it does not in general.
     
   The outcomes of quantum measurements are random variables, and the value of the loss function between an estimate and the true density matrix is also a random variable.
   Thus, in order to evaluate the precision of a general estimator $\rho^{\mathrm{est}}$ (not the estimate) for the true density matrix, we use the statistical expectation value of the loss function, called an \emph{expected loss} (sometimes called a risk function)\footnote{There are also different approaches to evaluating the precision of estimators, including error probabilities \cite{Sugiyama2011}, region estimators \cite{AudenaertScheel2009, Robin2012}.}. 
   The explicit form is given by
   \begin{eqnarray}
     \bar{\Delta}_{N}(\rho^{\mathrm{est}}| \rho)&:=&\sum_{x^N \in \mathcal{X}^N} p(x^N | \rho) \Delta (\rho^{\mathrm{est}}_{N}(x^{N}), \rho).     
   \end{eqnarray}
   The value of the expected loss depends on the choice of the estimator as well as the true density matrix.
   The latter is of course unknown in an experiment, and one way to eliminate its dependence is to average over all possible true states
   \begin{eqnarray}
      \bar{\Delta}^{\mathrm{ave}}_{N}( \rho^{\mathrm{est}}) &:=& \int_{\rho \in \mathcal{S}} \!\!\!\!\!\!\! d\mu(\rho) \bar{\Delta}_{N}( \rho^{\mathrm{est}}| \rho), \label{eq:ave}
   \end{eqnarray}
   where $\mu$ is a probability measure on $\mathcal{S}$. 
   The purpose of this paper is to clarify the behavior of expected losses for true states close to or on the boundary of $B$, so we focus not on average but pointwise expected losses for those states.
     
  Let us assume that $\| \bm{s} \| < 1$.
  For any unbiased estimator $\bm{s}^{\mathrm{est}}$ and any positive semidefinite matrix $H_{\bm{s}}$, the inequality
  \begin{eqnarray}
     &\bar{\Delta}_{N}& (\bm{s}^{\mathrm{est}} | \bm{s}) \notag \\
     &:= &\sum_{x^N \in \mathcal{X}^N} p(x^N |\bm{s}) [ \bm{s}^{\mathrm{est}}_{N}(x^N) - \bm{s} ]^T H_{\bm{s}} [ \bm{s}^{\mathrm{est}}_{N}(x^N) - \bm{s} ] \nonumber \\  
     &\phantom{:}\ge& \frac{1}{N} \tr [H_{\bm{s}} F_{\bm{s}}^{-1} ] \label{eq:gCRineq}
  \end{eqnarray}
  holds, where
  \begin{eqnarray}
     F_{\bm{s}} & :=& \sum_{x \in \mathcal{X}} \frac{\nabla_{\bm{s}} p(x |\bm{s}) \nabla_{\bm{s}}^{T} p(x |\bm{s})}{p(x |\bm{s})}, \\
                       & \phantom{:} =& \sum_{x \in \mathcal{X}} \frac{ \bm{w}_{x} \bm{w}_{x}^{T} }{ v_{x} + \bm{w}_{x} \cdot \bm{s} } \label{eq: Fisher}
  \end{eqnarray}
  is called the Fisher matrix and $\tr$ denotes the trace operation with respect to the parameter space $\mathbbm{R}^3$.
  Equation (\ref{eq:gCRineq}) is called the Cram\'{e}r-Rao inequality, and it holds not only for one-qubit state tomography, but also for arbitrary finite dimensional parameter estimation problems under some regularity condition \cite{RaoText1973}.
  The matrix $F_{\bm{s}}$ is a $3 \times 3$ positive semidefinite matrix for $\bm{s} \in \mathbbm{R}^3$.   
  It is known that a maximum likelihood estimator asymptotically achieves the equality of Eq.(\ref{eq:gCRineq}) \cite{RaoText1973}. 
  From the explicit formulas for the squared Hilbert-Schmidt distance and infidelity in Eqs. (\ref{eq:HSdistance2_para}) and (\ref{eq:Infidelity_para}), we have 
  \begin{eqnarray}
     \Delta^{\mathrm{HS}} (\bm{s} , \bm{s}^{\prime} ) &=& (\bm{s}^{\prime} - \bm{s} )^{T}\frac{1}{4}I(\bm{s}^{\prime} - \bm{s} ), \\
     \Delta^{\mathrm{IF}} (\bm{s} , \bm{s}^{\prime} ) &=& (\bm{s}^{\prime} - \bm{s} )^{T}\frac{1}{4}\Bigl(I + \frac{\bm{s}\bm{s}^{T}}{1-\| \bm{s}\|^2} \Bigr)(\bm{s}^{\prime} - \bm{s} ) \notag \\
      & & + O(\|\bm{s}^{\prime} - \bm{s} \|^3), \label{eq:infidelity_2nd}
  \end{eqnarray}
  where $I$ is the identity matrix on $\mathbbm{R}^3$.
  Therefore when we use the Hilbert-Schmidt distance as our loss function, we substitute $H_{\bm{s}}$ in Eq. (\ref{eq:gCRineq}) by $H^{\mathrm{HS}}_{\bm{s}} := \frac{1}{4}I$. 
  On the other hand, when our loss function is the infidelity, we must use $H^{\mathrm{IF}}_{\bm{s}} := \frac{1}{4}\Bigl(I + \frac{\bm{s}\bm{s}^{T}}{1-\| \bm{s}\|^2} \Bigr)$.
  These two matrices $H_{\bm{s}}^{\mathrm{HS}}$ and $H_{\bm{s}}^{\mathrm{IF}}$ are half of the Hesse matrices for $\Delta^{\mathrm{HS}}$ and $\Delta^{\mathrm{IF}}$, respectively. 
  
  The Cram\'{e}r-Rao inequality is often used to evaluate the estimation errors of a maximum likelihood estimator, but there are problems applying the bound to evaluating the expected losses for finite data sets. 
   The inequality holds for unbiased estimators, which maximum likelihood estimators are asymptotically.  However, they are not unbiased for finite $N$, because of the existence of the boundary in the parameter space.  
  It has been shown numerically that for values of $N$ in typical experiments the Cram\'{e}r-Rao bound cannot be applied \cite{Sugiyama2012_1}. 
  Hence, we are motivated to investigate the behavior of expected losses in parameter spaces with boundaries for finite data sets.
  We undertake this investigation for one-qubit in the next section.

   \section{Theoretical analysis}\label{sec:theory}
   
      In this section, we derive a function which approximates the expected losses of the squared Hilbert-Schmidt distance and infidelity for finite data sets. 
   
      \subsection{Two approximations}
        
         In general, the explicit form of expected losses with finite data sets is extremely complicated.
         In this paper, we try to derive not the exact form but a simpler function which reproduces the behavior of the true function accurately enough to help us understand the boundary effect. 
         In order to accomplish this, we introduce two approximations.
         First, we approximate the multinomial distribution generated by successive trials by a Gaussian distribution.
         Second, we approximate the spherical boundary by a plane tangent to its boundary. 
         
         %
          From the central limit theorem, we can readily prove that the distribution of a linear estimator $\bm{s}^{\mathrm{li}}$ converges to a Gaussian distribution with mean $\bm{s}$ and covariance matrix $F_{\bm{s}}^{-1}$.
           For finite $N$, we approximate the true probability distribution by the Gaussian distribution
          \begin{eqnarray}
             &p_{G}(\bm{s}^{\mathrm{li}}_{N} | \bm{s}) := & \notag \\
             &\frac{N^{3/2}}{(2\pi)^{3/2}\sqrt{\det F_{\bm{s}}^{-1}}}& \exp \bigl[ - \frac{N}{2} (\bm{s}_{N}^{\mathrm{li}} - \bm{s}) \cdot F_{\bm{s}} (\bm{s}_{N}^{\mathrm{li}} - \bm{s})  \bigr] .
          \end{eqnarray} 
         We will refer to this as the Gaussian distribution approximation (GDA).
         Because the approximation of the multinomial distribution by the GDA becomes better as each outcome probability grows sufficiently larger than 0, the expected losses under the GDA should be closer to the true expected losses the farther the true Bloch vector is from alignment with the axes in the Bloch sphere defined by the measurement.
         
         For a one-qubit system, the boundary between the physical and unphysical regions of the state space is a sphere with unit radius.
         Despite its simplicity, it is difficult to derive the explicit formula of a maximum likelihood estimator even in this case.
         Indeed, this is a major contributor to the general complexity of the expected loss behavior in quantum tomography.
         We therefore choose the simplest possible way to approximate the boundary, namely by replacing it with a plane in the state space.
         Suppose that the true Bloch vector is $\bm{s} \in B$.
         The boundary of the Bloch ball, $\partial B$, is represented as 
         \begin{eqnarray}
            \partial B := \{ \bm{s}^{\prime} \in \mathbbm{R}^3 |\ \| \bm{s}^{\prime} \| = 1 \}.
         \end{eqnarray}  
         We approximate this by the tangent plane to the sphere at the point $\bm{e}_{\bm{s}} := \bm{s} / \| \bm{s} \| $,  represented as 
         \begin{eqnarray}
            \partial D_{\bm{s}} := \{ \bm{s}^{\prime} \in \mathbbm{R}^{3} |\ \bm{s} \cdot (\bm{s}^{\prime} - \bm{e}_{\bm{s}} ) = 0 \},
         \end{eqnarray}
         and so the approximated parameter space is represented as
         \begin{eqnarray}
             D_{\bm{s}} = \{  \bm{s}^{\prime} \in \mathbbm{R}^{3} | \ \bm{s} \cdot (\bm{s}^{\prime} - \bm{e}_{\bm{s}} ) \le 0 \}.
         \end{eqnarray} 
         We will refer to this as the linear boundary approximation (LBA).
         The LBA is a specific case of tangent cone methods in statistical estimation theory which have been developed and used for analyzing models with constrained parameters in classical statistical estimation theory \cite{Chernoff1954, SelfLiang1987}.   
         It is known that the distribution of a maximum likelihood estimator in a constrained parameter estimation problem converges to the Gaussian distribution with a boundary approximated by a tangent cone \cite{SelfLiang1987}.
         Therefore it is guaranteed that the expected losses approximated by the GDA and LBA converge to their true values in the limit of infinite data.

      %
      %
       \begin{table*}[htbp]
          \caption{\label{table:TypicalN} List of the true Bloch vectors under consideration (in spherical coordinates), and numerical values of $N^{*}$ (rounded down, when possible).}
          \begin{ruledtabular}
                \begin{tabular}{c|ccccc}
                   $(r, \theta , \phi)$ & $(0.9, 0, 0)$ & $(0.9, \pi /4 , \pi / 4)$ & $(0.99, 0, 0)$ & $(0.99, \pi /4 , \pi / 4)$ & $(1, \pi /4 , \pi / 4)$ \\
                   Panels                   & (EIF-1) & (EIF-2) & (EIF-3) & (EHS-1), (EIF-4) & (EHS-2), Figure \ref{fig:EIF_pure}. \\
                   \hline
                   $N^{*}$                   & 114 & 417 & 1194 & 37947 & $\infty$ \\                           
             \end{tabular}
          \end{ruledtabular}
       \end{table*}               
         
      \subsection{Approximated maximum likelihood estimator}
      
          In \cite{SelfLiang1987}, it is proved that the distribution of a maximum likelihood estimator in a constrained parameter estimation problem converges to the distribution of the following vector
          \begin{eqnarray}
             \tilde{\bm{s}}_{N}^{\mathrm{ml}} := \mathrm{argmin}_{\bm{s}^{\prime} \in D_{\bm{s}}} (\bm{s}^{\mathrm{li}}_{N} - \bm{s}^{\prime}) \cdot F_{\bm{s}} (\bm{s}^{\mathrm{li}}_{N} - \bm{s}^{\prime}).
          \end{eqnarray}
          By using the Lagrange multiplier method, we can derive the approximated maximum likelihood estimates as       
          \begin{eqnarray}
             \tilde{\bm{s}}^{\mathrm{ml}}_{N} = 
             \left\{
                \begin{array}{lc}
                   \bm{s}^{\mathrm{li}}_{N} & (\bm{s}^{\mathrm{li}}_{N} \in D_{\bm{s}}) \\
                   \bm{s}^{\mathrm{li}}_{N} - \frac{\bm{e}_{\bm{s}} \cdot \bm{s}^{\mathrm{li}}_{N} - 1}{\bm{e}_{\bm{s}} \cdot F_{\bm{s}}^{-1} \bm{e}_{\bm{s}} } F_{\bm{s}}^{-1} \bm{e}_{\bm{s}}  & (\bm{s}^{\mathrm{li}}_{N} \notin D_{\bm{s}})
                \end{array}
             \right. .
          \end{eqnarray}
          We note that $\tilde{\bm{s}}^{\mathrm{ml}}_{N}$ depends on the true parameter $\bm{s}$, and so by definition it is not an estimator -- it is a vector introduced for the purpose of approximating expected losses of a maximum likelihood estimator.  Intuitively, it takes the value of the linear estimate if that estimate is physical, and if it is unphysical a correction vector is added to bring it back within the physical region.
            
      \subsection{Expected squared Hilbert-Schmidt distance}\label{subsec:Theory-EHS}

        From a straightforward calculation using formulas for Gaussian integrals, we can derive the approximate expected squared Hilbert-Schmidt distance.
        \begin{eqnarray}
           \bar{\Delta}^{\mathrm{HS}}_{N}(\tilde{\bm{s}}^{\mathrm{ml}} | \bm{s}) 
              = & \frac{1}{4} \Bigl( \tr [F_{\bm{s}}^{-1}] -\frac{1}{2} \frac{ \bm{e}_{ \bm{s} } \cdot F_{\bm{s}}^{-2} \bm{e}_{ \bm{s} } }{ \bm{e}_{ \bm{s} } \cdot F_{\bm{s}}^{-1} \bm{e}_{ \bm{s} } }  \mathrm{erfc} \Bigl[  \sqrt{\frac{N}{N^{*}}} \Bigr]  \Bigr) \frac{1}{N} \nonumber \\          
               & - \frac{1}{4} \frac{1-\|  \bm{s} \|}{\sqrt{2\pi \bm{e}_{\bm{s}} \cdot F_{\bm{s}}^{-1} \bm{e}_{\bm{s}}}} \frac{ \bm{e}_{ \bm{s} } \cdot F_{\bm{s}}^{-2}  \bm{e}_{ \bm{s} } }{\bm{e}_{ \bm{s} } \cdot F_{\bm{s}}^{-1} \bm{e}_{ \bm{s} } }  \frac{e^{ - N / N^{*} }}{\sqrt{N}} \nonumber \\
               & +  \frac{1}{8} (1 - \| \bm{s}\|^2 )  \frac{\bm{e}_{ \bm{s} } \cdot  F_{\bm{s}}^{-2} \bm{e}_{ \bm{s} } }{ (\bm{e}_{ \bm{s} } \cdot F_{\bm{s}}^{-1}\bm{e}_{ \bm{s} } )^2 } \mathrm{erfc} \Bigl[ \sqrt{\frac{N}{N^{*}}}  \Bigr], \label{eq:EHS_erfc_mixed}
           \end{eqnarray}
           where 
           \begin{eqnarray}
               \mathrm{erfc}[a] := \frac{2}{\sqrt{\pi}} \int_{a}^{\infty} \! \! dt \ e^{- t^{2} }
           \end{eqnarray}
           is the complementary error function and
           \begin{eqnarray}
              N^{*} := 2\frac{\bm{e}_{\bm{s}} \cdot F_{\bm{s}}^{-1} \bm{e}_{\bm{s}}}{(1 - \| \bm{s} \|)^2} \label{eq:Nstar}
           \end{eqnarray}
           is a typical scale for the number of trials.
           By using the Cram\'{e}r-Rao inequality, Eq. (\ref{eq:gCRineq}), we can prove that $\bm{e}_{\bm{s}} \cdot F_{\bm{s}}^{-1} \bm{e}_{\bm{s}} / N$ is the variance of the linear estimates $\bm{s}^{\mathrm{li}}_{N}$ in the $\bm{e}_{\bm{s}}$ direction of the Bloch sphere.
           When $N$ is sufficiently large, most of the distribution of linear estimates is included in $D_{\bm{s}}$ and the effect of the boundary becomes negligible. 
           Roughly speaking, this condition is represented as $\bm{e}_{\bm{s}} \cdot F_{\bm{s}}^{-1} \bm{e}_{\bm{s}} / N \ll (1 - \| \bm{s} \|)^2$, where the right hand side is the squared Euclidean distance between $\bm{s}$ and $D_{\bm{s}}$.  This can be rewritten as 
           \begin{eqnarray}
              N \gg \frac{\bm{e}_{\bm{s}} \cdot F_{\bm{s}}^{-1} \bm{e}_{\bm{s}}}{(1 - \| \bm{s} \|)^2} = \frac{1}{2} N^{*}.
            \end{eqnarray}
            We interpret $N^{*}$ as a reasonable benchmark for judging whether most of the distribution of the linear estimates is included in the physical region or not.
            The factor of $2$ in Eq. (\ref{eq:Nstar}) comes from the Gaussian integration, though in defining $N^\ast$ it is fairly arbitrary as it makes precise what we mean by `most' in the preceding sentence.
             Thus, in order to justify the use of the Cram\'{e}r-Rao bound for evaluating the estimation error, the number of measurement trials, $N$, must be larger than $N^{*}$. 
           
           When $\| \bm{s} \| <1$, in the limit of $N \to \infty$, $\mathrm{erfc}[\sqrt{N/N^{*}}]$ decreases exponentially fast.
           This can be readily shown by using the asymptotic expansion \cite{AbramowitzStegun1972},
           \begin{eqnarray}
              \mathrm{erfc}[a] \sim \frac{e^{-a^{2}}}{\sqrt{\pi} a} \Bigl( 1 + \sum_{m=1}^{\infty} (-1)^m \frac{1\cdot 3 \cdots (2m-1)}{(2a^{2})^m} \Bigr) .
           \end{eqnarray}
           Therefore we can see that the approximate expected squared Hilbert-Schmidt distance converges to the Cram\'{e}r-Rao bound.
           On the other hand, when $\| \bm{s} \| = 1$, the first and second terms disappear and we obtain
           \begin{eqnarray}
              \bar{\Delta}^{\mathrm{HS}}_{N}(\tilde{\bm{s}}^{\mathrm{ml}} | \bm{s}) 
               = \frac{1}{4} \frac{1}{N} \Bigl( \tr [F_{\bm{s}}^{-1}] -\frac{1}{2} \frac{\bm{e}_{\bm{s}} \cdot F_{\bm{s}}^{-2} \bm{e}_{\bm{s}} }{ \bm{e}_{\bm{s}} \cdot F_{ \bm{s} }^{-1} \bm{e}_{\bm{s}} } \Bigr),   \label{eq:EHS_erfc_pure}           
           \end{eqnarray}
           where we assumed that $F_{\bm{s}} < \infty$ for a Bloch vector $\bm{s}$ with $\| \bm{s} \| = 1$.
           This is smaller than the Cram\'{e}r-Rao bound, and this implies that when the true state is pure, a maximum likelihood estimator can break the Cram\'{e}r-Rao bound even in the asymptotic region.

          %
          %
          \subsection{Expected infidelity}\label{subsec:Theory-EIF}
      
            
           In order to analyze the expected infidelity, we take the Taylor expansion of the infidelity around the true Bloch vector $\bm{s}$ up to the second order.
           The explicit form is in Eq. (\ref{eq:infidelity_2nd}).
           Again, using formulas for Gaussian integrals we can derive the approximate expected infidelity.
           When $\| \bm{s} \| <1$,
         \begin{eqnarray}
            &\bar{\Delta}_{N}^{\mathrm{IF}}& (\tilde{\bm{s}}^{\mathrm{ml}} | \bm{s}) \notag \\
               &= &  \frac{1}{4}\Bigl(\tr [F_{\bm{s}}^{-1}] + \frac{\bm{s} \cdot F_{ \bm{s} }^{-1} \bm{s}}{1 - \| \bm{s}\|^2} \Bigr) \frac{1}{N} \Bigl(1 - \frac{1}{2}\mathrm{erfc} \Bigl[ \sqrt{\frac{N}{N^{*}}} \Bigr] \Bigr) \nonumber \\
                 &  &  -\frac{1}{4} \frac{1-\|\bm{s}\| }{ \sqrt{2\pi \bm{e}_{\bm{s}} \cdot F_{\bm{s}}^{-1} \bm{e}_{\bm{s}}} } \times \notag \\
                 &  &     \Bigl( \tr [F_{\bm{s}}^{-1}] - \tr [(Q_{\bm{s}} F_{\bm{s}} Q_{\bm{s}})^{-}] + \frac{ \bm{s} \cdot F_{\bm{s}}^{-1} \bm{s} }{1-\| \bm{s} \|^2 }  \Bigr) \frac{e^{-N/N^{*}}}{\sqrt{N}} \nonumber \\
                 &  & +\frac{1}{4} (1 - \| \bm{s} \|) \mathrm{erfc}\Bigl[ \sqrt{\frac{N}{N^{*}}} \Bigr] , \label{eq:EIF_erfc_mixed}
         \end{eqnarray} 
         where
         \begin{eqnarray}
            Q_{\bm{s}} := I - \bm{e}_{\bm{s}} \bm{e}_{\bm{s}}^{T}
         \end{eqnarray}
         is the projection matrix onto the subspace orthogonal to $\bm{s}$, and $A^{-}$  is the Moore-Penrose generalized inverse of a matrix $A$.
         From the argument above, we can see that the approximate expected infidelity converges to the Cram\'{e}r-Rao bound in the limit of large $N$.
         
         When $\| \bm{s} \|=1$, the infidelity is a 1st order function of $\bm{s}$, given by $\Delta^{\mathrm{IF}} (\bm{s}, \bm{s}^{\prime}) = \frac{1}{2}( 1 - \bm{s} \cdot \bm{s}^{\prime} )$, and there are no 2nd-order terms. 
          Consequently, the Hesse matrix of the infidelity $H^{\mathrm{IF}}_{\bm{s}}$ diverges at $\| \bm{s} \| =1$.
          Therefore we cannot apply the Cram\'{e}r-Rao inequality to the infidelity for pure states.
          By calculating the expectation value of the approximate estimator $\tilde{s}^{\mathrm{ml}}_{N}$, we can obtain 
         \begin{eqnarray}
            \bar{\Delta}_{N}^{\mathrm{IF}} (\tilde{\bm{s}}^{\mathrm{ml}} | \bm{s}) 
               = \frac{1}{2} \sqrt{ \frac{\bm{e}_{\bm{s}} \cdot F(\bm{s})^{-1} \bm{e}_{\bm{s}}}{2\pi} } \frac{1}{\sqrt{N}}. \label{eq:EIF_erfc_pure}
         \end{eqnarray}
                  
%

   %
   %
   \section{Numerical analysis}\label{sec:Numerical}

     We performed Monte Carlo simulations of one-qubit state tomography using three orthogonal projective (XYZ) measurements. 
     Our task is to estimate the density matrix of the one-qubit system, where the true state can be pure or mixed.
     We choose a maximum likelihood estimator, and we used a Newton-Raphson method to solve the (log-)likelihood equation with the completely mixed state $\bm{s}=\bm{0}$ as the initial point of the iteration.
     When the procedure returned a candidate point outside of the Bloch sphere, we chose the previous point (within the sphere) as the estimate.

      The POVM corresponding to three orthogonal projective measurements is given by 
      \begin{eqnarray} 
         \left\{  \frac{1}{3} | \alpha + \rangle \langle \alpha + | , \frac{1}{3} | \alpha - \rangle \langle \alpha - |  \right\}_{\alpha = 1, 2, 3}  ,
      \end{eqnarray}
      where $|\alpha \pm \rangle$ are the eigenstates of $\sigma_{\alpha}$ with eigenvalue $\pm 1$.
      The Fisher matrix and its inverse are given by
      \begin{equation}
         F_{\bm{s}} = \frac{1}{3} 
            \left(
               \begin{array}{ccc}
                  \frac{1}{1 - s_{1}^2} & 0 & 0 \\
                  0 & \frac{1}{1-s_{2}^2} & 0 \\
                  0 & 0 & \frac{1}{1 - s_{3}^2}
               \end{array}
            \right), \notag 
      \end{equation}
      \begin{equation}
          F_{\bm{s}}^{-1} = 3 
            \left(
               \begin{array}{ccc}
                  1 - s_{1}^2 & 0 & 0 \\
                  0 & 1-s_{2}^2 & 0 \\
                  0 & 0 & 1 - s_{3}^2
               \end{array}
            \right) \! .          \label{eq:Fisher-XYZ}
      \end{equation}

          In Secs. \ref{subsec:EHS-numerical} and \ref{subsec:EIF-numerical}, we show the plots for two loss functions: the squared Hilbert-Schmidt distance $\Delta^{\mathrm{HS}}$ and the infidelity $\Delta^{\mathrm{IF}}$.
     The pointwise expected losses $\bar{\Delta}_{N}(\bm{s}^{\mathrm{ml}} | \bm{s})$ and the approximated functions $\bar{\Delta}_{N} ( \tilde{\bm{s}}^{\mathrm{ml}} | \bm{s} )$ introduced in Sec. \ref{sec:theory} are compared, and the accuracy of those approximations are discussed.
     Table \ref{table:TypicalN} is a list of true Bloch vectors $\bm{s}$ for the figures shown in the following subsections, along with the numerical values of $N^{*}$ for each $\bm{s}$. 
     We chose two Bloch radii, $r=0.9$, $0.99$, and two set of angles $(\theta , \phi) = (0,0), (\pi / 4, \pi / 4)$ as the true Bloch vector $\bm{s}$.
     For a fixed $r$, the case with angles $(0,0)$ corresponds to one of the best case scenarios because the Bloch vector is along one measurement axis, while the $(\pi / 4, \pi /4)$ case corresponds to a worst case scenarios because the Bloch vector is equidistant from all the measurement axes.
     The explicit form of $N^{*}$ for the Fisher matrix in Eq. (\ref{eq:Fisher-XYZ}) is
     \begin{eqnarray}
        N^{*} = 6 \Big( \frac{1+\|\bm{s}\|}{1 - \|\bm{s} \|} + 2 \frac{ (s_{1} s_{2})^{2} + (s_{2} s_{3})^{2} + (s_{3} s_{1})^{2} }{\| \bm{s} \|^2 (1 - \| \bm{s}\|)^2 } \Bigr). \label{eq:N-XYZ}
     \end{eqnarray} 
     There are two terms which contribute to the divergence at $\| \bm{s} \| = 1$, and near this value the first term behaves as $O((1-\| \bm{s} \|)^{-1})$, while the second does as $O((1-\| \bm{s} \|)^{-2})$. 
     When the true Bloch vector is along one of the measurement axes, the second term in Eq. (\ref{eq:N-XYZ}) disappears.
     For example, if $\bm{s}= (r,0,0)$, we obtain $N^{*} = 6 \frac{1+r}{1-r} \sim \frac{12}{1-r}$ as $r\to 1$. 
     On the other hand, when the true Bloch vector does not lie along any measurement axis, the second term remains.
     For example, if $\bm{s} = (r , \pi / 4 , \pi / 4)$, we obtain $N^{*} = 6\Bigl( \frac{1+r}{1-r} + \frac{5}{8}\frac{1}{r^{2}(1-r)^2} \Bigr) \sim \frac{15}{4}\frac{1}{(1-r)^2}$. 
     Therefore $N^{*}$ for a true Bloch vector whose direction is along one of the measurement axes becomes smaller than that for a true Bloch vector whose direction is not. 
     This difference caused by the alignment of measurement axes becomes larger as the purity of $\rho(\bm{s})$ becomes higher.  
 
       \begin{figure}[t!]
          \begin{center}
             \includegraphics[width =0.9\linewidth]{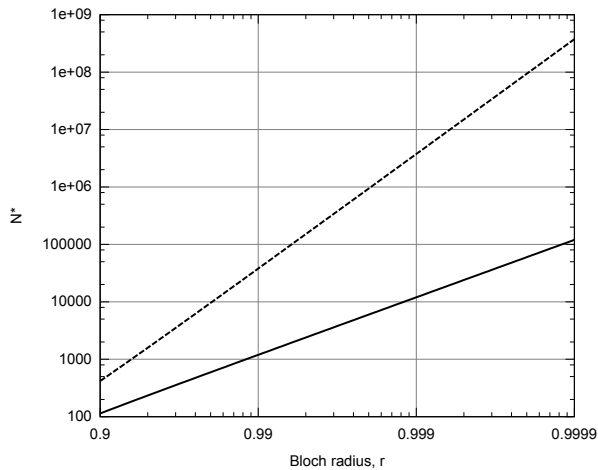}%
          \end{center}
          \caption{\label{fig:TypN}  Bloch radius dependency of $N^{*}$ for standard quantum state tomography, given in Eq. (\ref{eq:N-XYZ}). 
          The solid line is for states $\bm{s}$ given by $(r, 0, 0)$, and the dashed line is for those given by $(r, \pi / 4 , \pi / 4)$. }
       \end{figure}     

       \begin{figure*}[t!]
          \begin{center}
             \includegraphics[width =0.9\linewidth]{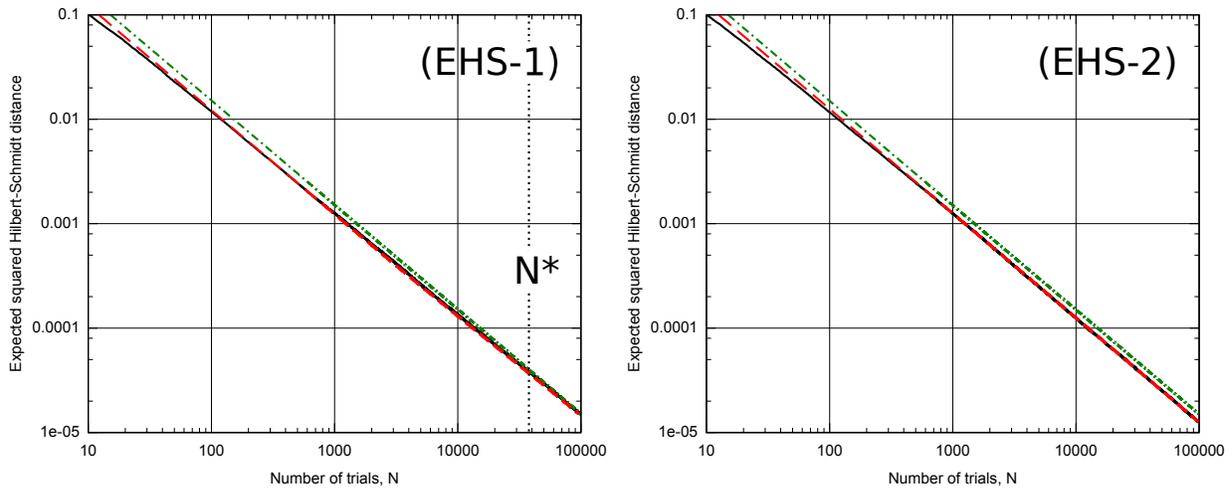}%
          \end{center}
          \caption{\label{fig:EHS} Pointwise expected squared Hilbert-Schmidt distance $\bar{\Delta}_{N}^{\mathrm{HS}}$ plotted against the number of measurement trials $N$: (EHS-1) and (EHS-2) are for the true Bloch vector $\bm{s}$ given by $(r, \theta , \phi) = (0.99, \pi / 4 , \pi / 4)$ and $(1, \pi / 4 , \pi / 4)$, respectively. The number of sequences used for the calculation of the statistical expectation values is 10 000. }
       \end{figure*}        
            
     The terms caused by the boundary in Eqs. (\ref{eq:EHS_erfc_mixed}), (\ref{eq:EHS_erfc_pure}), (\ref{eq:EIF_erfc_mixed}), (\ref{eq:EIF_erfc_pure}) start to decrease exponentially fast after $N$ becomes larger than $N^{*}$.
     We expect that the simulated and approximated plots start to converge to the Cram\'{e}r-Rao bound after $N$ becomes larger than $N^{*}$.        
     In all figures, the line styles are as follows: 
       a solid (black) line for the numerically simulated expected loss $\bar{\Delta}_{N}(\bm{s}^{\mathrm{ml}} | \bm{s})$, 
        a dashed (red) line for the approximate expected loss $\bar{\Delta}_{N} ( \tilde{\bm{s}}^{\mathrm{ml}} | \bm{s} )$ given in Eqs (\ref{eq:EHS_erfc_mixed}), (\ref{eq:EHS_erfc_pure}), (\ref{eq:EIF_erfc_mixed}), (\ref{eq:EIF_erfc_pure}),
        a chain (green) line for the Cram\'{e}r-Rao bound,
        and a dotted (black) vertical line for $N^{*}$.

     %
     %
      \subsection{Expected squared Hilbert-Schmidt distance}\label{subsec:EHS-numerical}\label{subsec:Numerical-EHS}

      The Cram\'{e}r-Rao bound of the expected squared Hilbert-Schmidt distance is given by 
      \begin{eqnarray}
         \frac{ \tr [H_{\bm{s}}^{\mathrm{HS}} F_{\bm{s}}^{-1}] }{ N } = \frac{3}{4} (3 - \| \bm{s} \|^2 ) \frac{1}{N}. \label{eq:EHS_CRB}
      \end{eqnarray}
        Figure \ref{fig:EHS} shows the pointwise expected squared Hilbert-Schmidt distance $\bar{\Delta}_{N}^{\mathrm{HS}}$ plotted against the number of trials $N$ (the horizontal and vertical axes are both logarithmic scale). 
        The panels (EHS-1) and (EHS-2) are for the true Bloch vector $\bm{s}$ given by $(r, \theta , \phi) = (0.99, \pi / 4 , \pi / 4)$ and $ (r, \theta , \phi) = (1, \pi / 4 , \pi / 4)$, respectively, so that the former is (slightly) mixed, while the latter is pure.  
        The panel (EHS-1) shows that our approximation in Eq. (\ref{eq:EHS_erfc_mixed}) converges to the simulated plot, and both the simulated and approximated plots converge to the Cram\'{e}r-Rao bound of Eq. (\ref{eq:EHS_CRB}) as $N$ becomes large.
        The same behavior is observed for other mixed true states.
        On the other hand, panel (EHS-2) shows a different behavior; our approximation in Eq. (\ref{eq:EHS_erfc_pure}) converges to the simulated plot, but the simulated and approximated plots do not converge to the Cram\'{e}r-Rao bound.
        This indicates that for pure states, our approximation better captures the behavior of the expected loss than does the Cram\'{e}r-Rao bound. 
         As mentioned around Eq. (\ref{eq:EHS_erfc_pure}), the reason for this is that the center of the distribution of the linear estimates for a pure state will always be on the boundary of the Bloch sphere, so that about a half of the distribution will always be in the unphysical region.  
         This prohibits a maximum likelihood estimator from ever converging to the Cram\'{e}r-Rao bound.

          \begin{figure*}[t!]
             \begin{center}
                \includegraphics[width =0.9\linewidth]{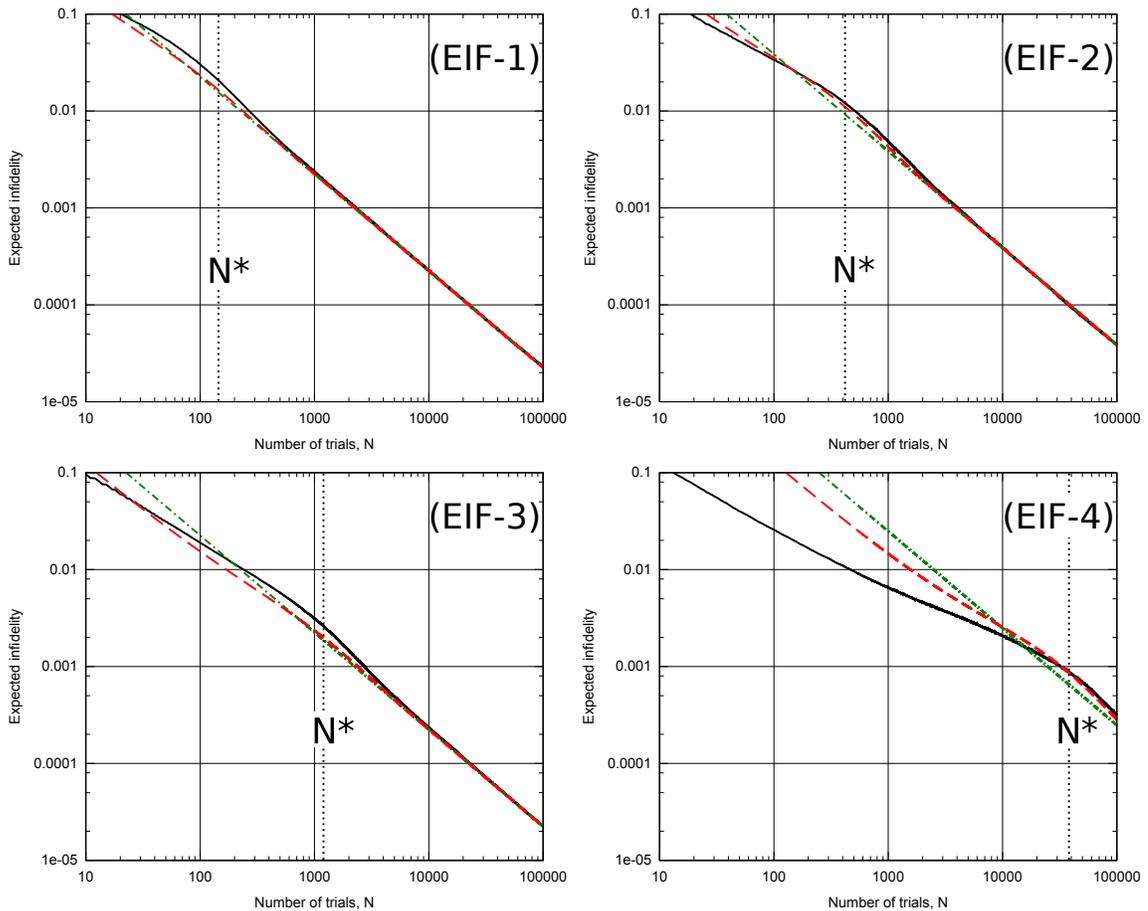}%
             \end{center}
             \caption{\label{fig:EIF_mixed} Pointwise expected infidelity $\bar{\Delta}_{N}^{\mathrm{IF}}$ plotted against the number of measurement trials $N$: (EIF-1), (EIF-2), (EIF-3), (EIF-4) are for the true Bloch vector $\bm{s}$ given by $(r, \theta , \phi) = (0.9, 0, 0), (0.9, \pi / 4 , \pi / 4), (0.99, 0, 0)$, and $(0.99, \pi / 4 , \pi / 4)$, respectively.  The number of sequences used for the calculation of the statistical expectation values is 10000.}
          \end{figure*}     
          
          \begin{figure}[t!]
             \begin{center}
                \includegraphics[width =0.9\linewidth]{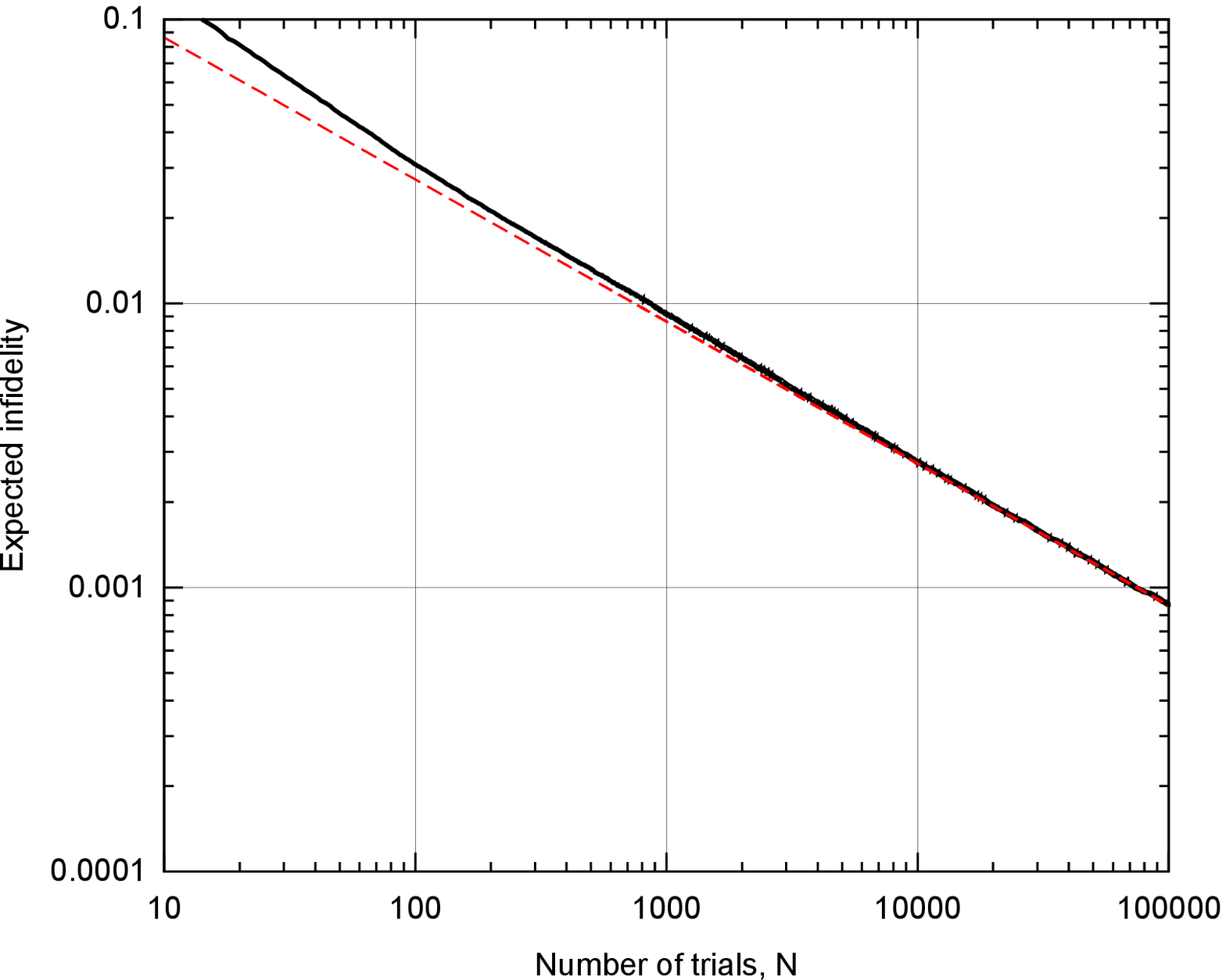}
             \end{center}
             \caption{\label{fig:EIF_pure} Pointwise expected infidelity $\bar{\Delta}_{N}^{\mathrm{IF}}$ plotted against the number of measurement trials $N$ for the true Bloch vector $\bm{s}$ given by $(r, \theta , \phi) = (1, \pi / 4 , \pi / 4)$. The number of sequences used for the calculation of statistical expectation values is 10 000.}
          \end{figure}

    %
    %
    \subsection{Expected infidelity }\label{subsec:EIF-numerical}\label{Numerical-EIF}
       
       The infidelity is a nonlinear function of the states, and we must approximate the Cram\'{e}r-Rao bound in this case; doing so up to second order gives
       \begin{eqnarray}
          \frac{ \tr [H_{\bm{s}}^{\mathrm{IF}} F_{\bm{s}}^{-1}] }{N} = \frac{3}{4} \Bigl( 3 + 2\frac{ (s_{1} s_{2})^{2} + (s_{2} s_{3})^{2} + (s_{3} s_{1})^{2} }{ 1 - \| \bm{s}\|^2 } \Bigr) \frac{1}{N}. \label{eq:EIF_CRB}
      \end{eqnarray}
       Figure \ref{fig:EIF_mixed} shows the pointwise expected infidelity $\bar{\Delta}_{N}^{\mathrm{IF}}$ plotted against the number of measurement trials $N$: (EIF-1), (EIF-2), (EIF-3), (EIF-4) are for the true Bloch vector $\bm{s}$ given by $(r, \theta , \phi) = (0.9, 0, 0), (0.9, \pi / 4 , \pi / 4), (0.99, 0, 0)$, and $(0.99, \pi / 4 , \pi / 4)$, respectively.
          Thus panels (EIF-1,2) and panels (EIF-3,4) ) are for true states with the same purity.
          Panels (EIF-1) and (EIF-3) are for the case that one of the measurement axes coincides with the direction of the true Bloch vector, while panels (EIF-2) and (EIF-4) are for the case that all of the measurement axes are as far as possible from the true Bloch vector.
          Figure \ref{fig:EIF_mixed} shows that $N^{*}$ is a good benchmark for the number of trials required for the simulated plot to start to converge to the Cram\'{e}r-Rao bound, and so we can say that in order to justify the use of the asymptotic theory, $N$ must be larger than $N^{*}$.
          Figure \ref{fig:EIF_mixed} indicates that the angle dependency of the expected infidelity becomes larger as the purity becomes higher.
          When the true state is far from all measurement axes, the accuracy of our approximation is higher than that of the Cram\'{e}r-Rao bound.  
          For $N$ smaller than about 10 000 (the `low $N$ region'), the accuracy of our approximation is low (though still higher than that of the Cramer-Rao bound).
          We believe that the main reason for our approximation's poor performance in this low $N$ region is the second order approximation of the infidelity, and that higher orders would improve the accuracy here.  However, in the high $N$ region the approximation can be seen to capture the behavior of the curve far better than the Cram\'{e}r-Rao bound.

          Figure \ref{fig:EIF_pure} shows the pointwise expected infidelity $\bar{\Delta}_{N}^{\mathrm{IF}}$ against the number of measurement trials $N$ for the true Bloch vector $\bm{s}$ given by $(r, \theta , \phi) = (1, \pi / 4 , \pi / 4)$.   
          For pure true states, the expected infidelity decreases as $O(\sqrt{N})$, and Fig. \ref{fig:EIF_pure} shows that the expected infidelity converges to the approximate function.  

%
%
\section{Conclusions}\label{sec:Conclusion}

   In this paper, we analyzed expected losses in one-qubit state tomography for finite data sets.
   We derived an explicit formula of the expected squared Hilbert-Schmidt distance and the expected infidelity between a tomographic maximum likelihood estimate and the true state under two approximations: a Gaussian distribution matched to the moments of the asymptotic multinomial distribution, and a linearization of the parameter space boundary imposed by the positivity of quantum states.
   We performed Monte Carlo simulations of one-qubit state tomography and evaluated the accuracy of the approximation formulas by comparing them to the numerical results.
   The numerical comparison shows that our approximation reproduces the behavior in the nonasymptotic regime much better than the asymptotic theory, and the typical number of measurement trials derived from the approximation is a reasonable threshold after which the expected loss starts to converge to the asymptotic behavior.
  

%
%
\section*{Acknowledgments}

T. S. would like to thank R. Blume-Kohout and C. Ferrie for their correspondence, as well as F. Tanaka for helpful discussion on mathematical statistics.  Additionally, P. S. T. would like to thank D. Mahler, L. Rozema and A. Steinberg for discussions pointing out the importance of this problem.
This work was supported by JSPS Research Fellowships for Young Scientists (22-7564) and Project for Developing Innovation Systems of the Ministry of Education, Culture, Sports, Science and Technology (MEXT), Japan.

%
%

%
%
\section*{References}

\end{document}